\documentclass[prl,twocolumn,showpacs]{revtex4}
\usepackage{graphicx}
\begin{document}
\title{Feshbach resonances in rubidium 87: Precision measurement and analysis}
\author{A. Marte$^1$, T. Volz$^1$, J. Schuster$^1$, S. D{\"u}rr$^1$, G. Rempe$^1$, E. G.\ M.\ van Kempen$^2$, and
B. J.\ Verhaar$^2$} 
\affiliation{$^1$Max-Planck-Institut f{\"u}r Quantenoptik, Hans-Kopfermann-Str.\ 1, 85748 Garching, Germany\\
$^2$Eindhoven University of Technology, P.O.\ Box 513, 5600MB Eindhoven, The Netherlands}
\date{\today}
\hyphenation{Fesh-bach}
\begin{abstract}
More than 40 Feshbach resonances in rubidium 87 are observed in the magnetic field range between 0.5 and 1260 gauss for various spin mixtures in the lower hyperfine ground state. The Feshbach resonances are observed by monitoring the atom loss, and their positions are determined with an accuracy of 30~mG. In a detailed analysis, the resonances are identified and an improved set of model parameters for the rubidium interatomic potential is deduced. The elastic width of the broadest resonance at 1007~G is predicted to be significantly larger than the magnetic field resolution of the apparatus. This demonstrates the potential for applications based on tuning the scattering length.
\end{abstract}
% 34.50.-s Scattering of atoms and molecules
% 03.75.Fi Phase coherent atomic ensembles; quantum condensation phenomena
% 32.80.Pj Optical cooling of atoms; trapping
% 34.20.Cf Interatomic potentials and forces
%
\maketitle
A Feshbach resonance is an exciting tool for controlling the atom-atom interaction in ultracold atomic gases. The elastic $s$-wave scattering length $a$ can be tuned over orders of magnitude simply by applying a magnetic field. Feshbach resonances have been observed in various alkali atoms \cite{inouye:98,courteille:98,roberts:98,vuletic:99,loftus:02,khayokovich:02,strecker:02}. They have been used to induce a controlled implosion of a Bose-Einstein condensate (BEC) \cite{donley:01}, to create a coherent superposition of an atomic BEC and a molecular state \cite{donley:02}, and to realize a bright soliton in a BEC \cite{khayokovich:02,strecker:02}. Future applications could include experiments with the Mott-insulator phase transition \cite{greiner:02}, a Tonks gas \cite{girardeau:01}, effects beyond the mean-field theory \cite{dalfovo:99}, and the creation of a molecular BEC \cite{kokkelmans:etal}. In addition, the binding energies of ro-vibrational molecular states close to the dissociation threshold can be determined with high accuracy from the position of Feshbach resonances leading to a precise knowledge of the interatomic potential \cite{chin:etal}. A Feshbach resonance showing up in elastic collisions is often accompanied by strong changes in the inelastic collision properties \cite{stenger:99,roberts:00}. This offers a strategy for searching for new Feshbach resonances by monitoring the resulting atom loss.

Surprisingly, no Feshbach resonance has been observed in $^{87}$Rb, which is the isotope used in most of today's BEC experiments. In 1995, a search with atoms in the $|f,m_f\rangle=|1,-1\rangle$ state was carried out, but no resonances were found \cite{newbury:95}. Meanwhile various experiments \cite{myatt:97,wynar:00,seto:00,roberts:01a} greatly improved the knowledge of the Rb interatomic potential. Recent models based on this \cite{vogels:97,kempen:02} are consistent with the observations of Ref.~\cite{newbury:95}. But these models predict four Feshbach resonances in the $|1,1\rangle$ state.

This letter reports the observation of more than 40 Feshbach resonances in $^{87}$Rb with most (but not all) atoms prepared in the $|1,1\rangle$ ground state. The theoretical model is extended to various spin mixtures and to bound states with rotational quantum number $l \leq 3$. Thus, all except one of the resonances can be clearly identified. Moreover, the measured position of one Feshbach resonance is used for an improved fit of the model parameters. The relative deviation between the predicted and observed positions is $1.6\times10^{-3}$ (rms). The observed loss might be due to two- or three-body inelastic collisions, but should be purely three-body for those resonances which involve only the $|1,1\rangle$ state. The broadest resonance at 1007~G offers the possibility to tune the scattering length and investigate its loss mechanism.

The experiment is performed with a new set-up similar to our previous one \cite{ernst:98}, but with all relevant components significantly improved. In particular, atoms are captured in a vapor-cell magneto-optical trap (MOT) at a loading rate of $7\times 10^{10}~{\rm s}^{-1}$. The atoms are transferred to a second MOT, in which $6\times 10^{9}$ atoms are accumulated by multiple transfer within 2~s. The atoms are then loaded into a Ioffe-Pritchard magnetic trap with a lifetime of 170~s. The measured bias-field drift of less than 1~mG/h illustrates the excellent stability of the magnetic trap. After 26~s of evaporative cooling, a BEC with up to $3.6 \times 10^6$ atoms is formed in the $|1,-1\rangle$ state.

The $|1,1\rangle$ state, in which the resonances are predicted, cannot be held in a magnetic trap. Therefore the atoms are now loaded into an optical dipole trap made of a single beam from an Yb:YAG laser at a wavelength of 1030~nm. A laser power of 45~mW is focused to a waist of 15~$\mu$m, resulting in measured trap frequencies of 930~Hz and 11~Hz, and an estimated trap depth of $k_B \times 20~\mu$K.

Once the atoms are in the optical trap, a radio-frequency field is used to transfer the atoms to the desired $|1,1\rangle$ state. With a Stern-Gerlach method, the fraction of atoms that end up in the $|1,1\rangle$ state is determined to be roughly 90\%. Almost all other atoms are in the $|1,0\rangle$ state. Next, a homogeneous magnetic field is applied in order to observe a Feshbach resonance. The field is created using the compensation coils of the magnetic trap which are in near-perfect Helmholtz configuration. Up to 1760 A of current are stabilized with a home-built servo to a few ppm. The magnetic field is held at a fixed value for typically 50~ms and then quickly switched off. Although many Feshbach resonances are rapidly crossed when the magnetic field is turned on or off, no significant loss of atoms is observed from these rapid crossings. After switching off the magnetic field, the atoms are released from the optical trap, and 14~ms later an absorption image of the expanded cloud is taken.

The search for Feshbach resonances was typically performed with a purely thermal cloud of $4 \times 10 ^6$ atoms in the optical trap at a temperature of 2~$\mu$K, corresponding to a peak density of $2\times 10^{14}$~cm$^{-3}$. 
The magnetic-field range between 0.5 and 1260~G was scanned and 43 Feshbach resonances were found. The magnetic-field values of these resonances $B_{exp}$ are listed in Tabs.~\ref{tab11+11} and \ref{tab-others}. The fields were calibrated with 30~mG precision using microwave spectroscopy in the vicinity of each resonance.

\begin{table}
\begin{tabular}{c|c|c|r|c|c|c}
$B_{\rm exp}$(G) & $B_{\rm th}$(G) & $d$(\%) & $\Delta$(mG) &
$l(f_1,f_2)v^{\prime},m_F$ & $F$ & $m_{f1}, m_{f2}$
\\ \hline
406.23  & 406.6  & 57 &       0.4 & 0(1,2)-4,\,2 &   & 0,2 \\
685.43  & 685.8  & 78 &        17 & 0(1,2)-4,\,2 &   & 1,1 \\
911.74  & 911.7  & 72 &       1.3 & 0(2,2)-5,\,2 & 4 &     \\
1007.34 & 1008.5 & 64 &       170 & 0(2,2)-5,\,2 & 2 &     \\
\hline
---     & 377.2  & ---& $\ll 0.1$ & 2(1,1)-2,\,0&   & -1,1 \\
---     & 395.0  & ---& $\ll 0.1$ & 2(1,1)-2,\,0 &   & 0,0  \\
856.85  & 857.6 &$<10$& $\ll 0.1$ & 2(1,1)-2,\,1 & 2 & 0,1 \\
\hline
---     & 249.1  & ---& $\ll 0.1$ & 2(1,2)-4,\,1 &   & -1,2 \\
306.94  & 306.2  & 34 & $\ll 0.1$ & 2(1,2)-4,\,0 &   & -1,1 \\
319.30  & 319.7  & 54 &   $< 0.1$ & 2(1,2)-4,\,2 &   & 0,2  \\
387.25  & 388.5  & 53 &   $< 0.1$ & 2(1,2)-4,\,1 &   & 0,1  \\
391.49  & 392.9  & 63 &       0.3 & 2(1,2)-4,\,3 & 3 & 1,2  \\
532.48  & 534.2  & 57 &   $< 0.1$ & 2(1,2)-4,\,0 &   & 0,0  \\
551.47  & 552.0  & 66 &       0.2 & 2(1,2)-4,\,2 &   & 1,1  \\
819.38  & 819.3  & 29 &   $< 0.1$ & 2(1,2)-4,\,1 &   & 1,0  \\
\hline
632.45  & 632.5  & 77 &       1.5 & 2(2,2)-5,\,4 & 4 & 2,2  \\
719.48  & 719.5  & 77 &       0.5 & 2(2,2)-5,\,3 & 4 & 1,2  \\
831.29  & 831.3  & 67 &       0.2 & 2(2,2)-5,\,2 & 4 & \\
930.02  & 930.9  & 78 &   $< 0.1$ & 2(2,2)-5,\,2 & 2 & \\
978.55  & 978.3  & 36 &   $< 0.1$ & 2(2,2)-5,\,1 & 4 & \\
1139.91 & 1140.9 & 10 & $\ll 0.1$ & 2(2,2)-5,\,1 & 2 & \\
---     & 1176.3 & ---& $\ll 0.1$ & 2(2,2)-5,\,0 & 4 & \\
\end{tabular}
\caption{
 \label{tab11+11}
Feshbach resonances in the $|1,1\rangle \otimes |1,1\rangle$ entrance channel. The experimentally observed positions $B_{\rm exp}$ are compared with the theoretical predictions $B_{\rm th}$ calculated at 2~$\mu$K. Also shown are the observed depths $d$, defined as the fraction of atoms lost during a 50-ms hold time. Note that the presence of atoms in spin states other than $|1,1\rangle$ prevents $d$ from reaching 100\%. For comparison, theoretical results for the elastic widths $\Delta$ of the resonances are listed. The last columns list the quantum numbers of the (quasi) bound state that gives rise to the resonance. Some very weak $l=2$ resonances could not be detected experimentally.}
\end{table}

\begin{figure}
\includegraphics{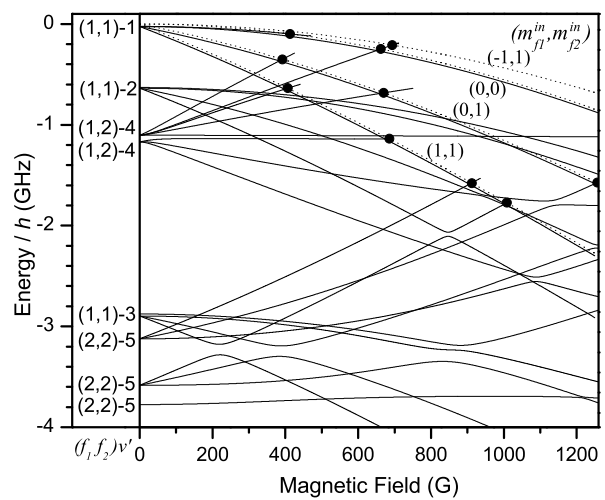}
\caption{
 \label{fig-theory}
$l=0$ Feshbach resonances in a coupled-channel calculation. Bound-state energies (solid lines) are shown as a function of magnetic field with quantum numbers $(f_1,f_2)v^{\prime}$ assigned at $B=0$. Additionally, dissociation threshold energies (dotted lines) are shown for four different entrance channels $(m_{f1}^{in},m_{f2}^{in})$. A Feshbach resonance ($\bullet$) occurs, when a bound state with quantum number $m_F$ crosses a dissociation threshold with the same $m_F$.}
\end{figure}

In order to analyze these results theoretically, both continuum and bound-state calculations are carried out using an accurate description of the atomic interaction between two Rb atoms \cite{kempen:02}. An introduction to the theory of Feshbach resonances can be found in Refs.~\cite{heinzen:99,burke:99}. The strong central part of the interaction, consisting of singlet and triplet potentials, conserves the orbital quantum numbers $l,m_l$ and the total-spin magnetic quantum number $m_F$ separately. (In bound-state spectroscopy, $l$ is often called $N$.) In the ultracold regime, with this part of the interaction included, Feshbach resonances are expected to occur only when an $l = 0$ bound state with a certain value of $m_F$ crosses the dissociation threshold of an entrance channel with the same $m_F$. Figure~\ref{fig-theory} shows these crossings in the $E$-$B$ plane as solid dots along various $(m^{in}_{f1},m^{in}_{f2})$ thresholds. The resonance fields obtained from these bound-state calculations are listed in Tabs.~\ref{tab11+11} and \ref{tab-others} ($l$ = 0 quantum number). The same field values also follow from the $B$-dependence of the elastic $S$-matrix element, as obtained in a continuum calculation.

In Fig.~\ref{fig-theory}, the bound states are labelled $(f_1,f_2)v^{\prime}$, with the vibrational quantum number $v^{\prime}$ = -1, -2, -3, \dots\ counting from the corresponding $(f_1,f_2)$ threshold. At $B=0$, the exchange interaction couples the atomic spins $f_1$, $f_2$ to a total molecular spin $F$ and causes a splitting between states with the same $(f_1,f_2)v^{\prime}$ but different $F$. In the presence of a strong external magnetic field, however, $m_{f1}$ and $m_{f2}$ become good quantum numbers instead of $F$, while $m_F$ is always a good quantum number. What constitutes a strong field in this sense depends on the size of the $F$-splitting at $B=0$. For small $|v^{\prime}|$, the $F$-splitting is hardly visible in Fig.~\ref{fig-theory}.

\begin{table}
\begin{tabular}{c|c|c||c|c|c}
$B_{\rm exp}$(G) & $B_{\rm th}$(G) & $l(m_{f1}^{in},m_{f2}^{in})$
&
$B_{\rm exp}$(G) & $B_{\rm th}$(G) & $l(m_{f1}^{in},m_{f2}^{in})$ \\
\hline
391.08  & 391.7  & 0(0,1)  & 825.11  & 825.4  & 3(0,1) \\
417.20  & 417.7  & 2(0,1)  & 965.96  & 966.0  & 1(0,1) \\
535.01  & 536.6  & 2(0,1)  & 1137.97 & 1135.5 & 1(0,1) \\
\cline{4-6}
548.60  & 550.7  & 2(0,1)  &  414.34  & 413.7  & 0(0,0) \\
669.19  & 670.7  & 0(0,1)  & 661.43  & 662.2  & 0(0,0) \\
802.94  & 805.0  & 2(0,1)  &  729.43  & 728.5  & 2(0,0) \\
821.04  & 821.7  & 2(0,1)  &   760.73  & 762.1  & 2(0,0) \\
840.95  & 841.0  & 2(0,1)  & 1167.14 & 1167.1 & 2(0,0) \\
981.54  & 981.7  & 2(0,1)  &  1208.69 & 1209.4 & 2(0,0) \\
1162.15 & 1162.5 & 2(0,1)  &  1252.68 & 1254.9 & 2(0,0) \\
\cline{4-6}
1236.73 & ---    & -(0,1)  &  692.75  & 693.6  & 0(-1,1) \\
1237.19 & 1238.1 & 2(0,1)  &  1216.32  & 1216.6 & 2(-1,1) \\
1256.96 & 1257.1 & 0(0,1)  &          &        &        \\
\end{tabular}
\caption{
 \label{tab-others}
Experimentally observed Feshbach resonances in other entrance channels $|1,m_{f1}^{in}\rangle \otimes |1,m_{f2}^{in}\rangle$.}
\end{table}

The much weaker spin-spin interaction $V_{ss}$ consists of the magnetic dipole-dipole interaction of the valence electrons together with a second-order spin-orbit term \cite{mizushima:etal}. Due to its tensor form it breaks the spatial spherical symmetry and allows a redistribution between the angular momenta of the spin and spatial degrees of freedom so that the sum $m_l+m_F$ is the only conserved quantum number. $V_{ss}$ admixes an $l$ = 0 component in otherwise pure $l$ = 2 bound states and therefore induces additional $s$-wave Feshbach resonances in the ultracold regime. Resonances of this type have previously been observed in Cs \cite{chin:etal}, but up to now never in $^{85}$Rb or $^{87}$Rb. The resulting resonance fields are again listed in the tables ($l$ = 2 resonances). For mixed species resonances, such as $|1,0\rangle \otimes |1,1\rangle$, the $l=1$ partial wave can be populated in the entrance channel, thus opening up the possibility to observe resonances due to $l=1$ or $l=3$ bound states.

The set of potential parameters used in this calculation is obtained as follows. The field value 911.74~G of one resonance is added to the set of eight experimental data already included in the analysis of Ref.~\cite{kempen:02}. This particular resonance was chosen, because the corresponding bound state is a pure triplet state and the previous experimental constraints on the singlet potential \cite{seto:00} are much stronger than those on the triplet potential. Then, a least-squares analysis is applied with the column A parameters of Tab.~I in Ref.~\cite{kempen:02} as starting values. This leads to an adjusted set of parameter values differing from the starting values by less than $1\sigma$; in atomic units: $C_6=4.707\times 10^3$, $C_8 = 5.73 \times 10^5$, $C_{10}=7.665\times 10^7$ (Ref.~\cite{marinescu:94}), $J = 0.486\times 10^{-2}$, $a_S(^{87}{\rm Rb})=+90.6$, $a_T(^{87}{\rm Rb})=+98.96$, $v_{DS}(^{87}{\rm Rb})=0.454$, and $v_{DT}(^{87}{\rm Rb})=0.4215$. Coupled-channel calculations \cite{heinzen:99,burke:99} based on these fine-tuned parameters then yield theoretical resonance fields $B_{th}$. The relative deviation of all these positions from the observations is $1.6\times10^{-3}$ (rms). Compared to the four resonance positions predicted in \cite{kempen:02}, this is an improvement of a factor of 6. Since only one resonance position was included in the fit, the excellent agreement with all other positions demonstrates the accuracy of the model. Note that the observed position of maximum loss might deviate from the pole of the elastic scattering length \cite{fedichev:etal}. However, such deviations are typically not larger than the elastic widths of the resonances; and they are small as discussed below.

Some resonances are so close together that they cannot be identified merely from their positions. In these cases, a Stern-Gerlach method was used to experimentally determine which $m_f$ states incurred the strongest atom loss. Thus the entrance channel could be identified, in particular when additionally varying the initial spin mixture. The entrance channel of the non-identified resonance at 1236.73~G was also determined with this method.

An interesting property of a Feshbach resonance -- besides its position -- is its strength. The strength of the elastic resonance is proportional to the field width $\Delta$ over which the scattering length has opposite sign \cite{burke:99}, as listed in Tab.~\ref{tab11+11}. The strength of the inelastic scattering properties is quantified by the loss rate. The dominant loss for the resonances in Tab.~\ref{tab11+11} arises from inelastic three-body collisions. This is because for the parameters of the experiment, single-body loss is negligible; and since the $|1,1\rangle$ entrance channel is the absolute ground state of atomic $^{87}$Rb, inelastic two-body collisions cannot occur \cite{stenger:99}. (This is not the case for the resonances in Tab.~\ref{tab-others}.) The three-body loss is characterized by the coefficient $K_3$ in the rate equation $\dot{N}= - K_3 \langle n^2 \rangle N$, where $N$ is the atom number and $n$ the density. The depth $d$ listed in Tab.~\ref{tab11+11} is a non-linear, yet monotonic function of $K_3$. In Tab.~\ref{tab11+11}, one finds a clear trend that stronger resonances with larger $\Delta$ cause faster loss, i.e.\ larger $d$. This is plausible, although the theory of three-body losses is not yet fully understood, especially in the vicinity of Feshbach resonances \cite{burke:99,fedichev:etal}.

Interestingly, the widths $\Delta$ of the $l=0$ resonances in Tab.~\ref{tab11+11} are much smaller than for other alkali atoms \cite{stenger:99,vuletic:99,roberts:01a,loftus:02}. This is due to the approximate phase equality of the waves reflected from the short-range singlet and triplet potentials, that is also responsible for other remarkable $^{87}$Rb phenomena: the coexistence of condensates in different hyperfine states \cite{myatt:97} and the smallness of the fountain-clock frequency shift \cite{kempen:02}.

All resonances display a nearly symmetric loss feature. A Gaussian is fit to the atom loss at a 50-ms hold time in order to extract the depth and the width of the atom loss. The Gaussian was chosen for convenience and because it fits well to the data. The obtained rms-widths are between 20~mG and 100~mG for almost all resonances. For small $d$, this width is identical to the width of the resonance in $K_3$ (if two-body loss is absent). For large $d$, however, the non-linear dependence of $d$ on $K_3$ broadens the observed width at a given hold time, somewhat similar to saturation broadening of spectral lines. In addition, the finite temperature of the cloud gives rise to a broadening of typically 20~mG and leads to a shift of similar size. The observed rms-width of 24(4)~mG at 965.96~G sets an experimental upper limit to thermal and technical broadening.

\begin{figure}
\includegraphics{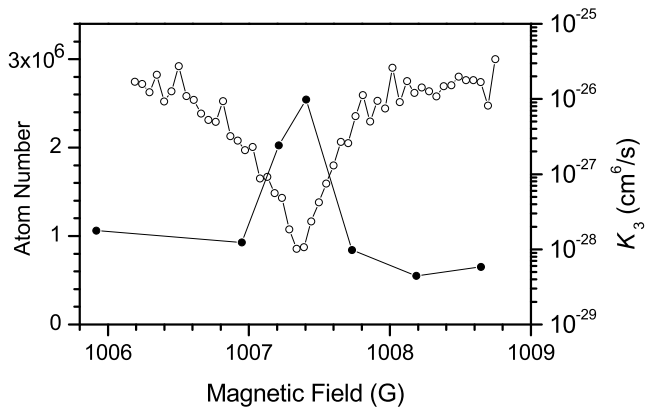}
\caption{
 \label{fig-exp}
Shape of the broadest Feshbach resonance. The number of atoms ($\circ$) remaining in the trap after a 50-ms hold time is displayed as a function of magnetic field. The three-body loss rate coefficient $K_3$ is also shown ($\bullet$).}
\end{figure}

The broadest resonance is centered at 1007~G (see Fig.~\ref{fig-exp}). The theoretical prediction for its elastic width, $\Delta = 170(30)$~mG, is large compared to the upper bound on the experimental broadening of 24(4)~mG (see above). This demonstrates the potential for a controlled variation of the scattering length with the present set-up.

For various field values, the decay of the atom number was also measured as a function of time. If one assumes that the loss can be described by a rate equation, three-body loss dominates (see above). Values of $K_3$ determined from a fit are shown in Fig.~\ref{fig-exp}. Note that the value $K_3=3.2(1.6)\times10^{-29}$~cm$^6$s$^{-1}$ obtained away from the resonance is consistent with the measured value $K_3=4.3(1.8)\times10^{-29}$~cm$^6$s$^{-1}$ for the $|1,-1\rangle$ state \cite{burt:97}. The absence of two-body loss makes this resonance an ideal candidate for testing theories of three-body loss \cite{fedichev:etal}. The values of $K_3$ in Fig.~\ref{fig-exp} are much smaller than those for $^{85}$Rb \cite{roberts:00}, where exciting experiments have been performed \cite{donley:01,donley:02}. Three-body loss will  therefore not be a substantial problem for applications.

All data obtained here fit well to a three-body decay, but the initial atom number extrapolated from the fit is a factor of up to 2 lower at the 1007~G resonance as compared to the off-resonance value. Since no data were taken for hold times shorter than 10~ms, this suggests the existence of an additional loss mechanism acting on much faster timescales. This could be molecule formation and dissociation during the first 10~ms of hold time \cite{claussen:etal}. Note, however, that here loss from a purely thermal cloud is observed, while loss from a BEC is discussed in Ref.~\cite{claussen:etal}.

To summarize, more than 40 Feshbach resonances have been observed in $^{87}$Rb. All except one were identified by theory. After including one of the observed resonances in fitting the model parameters, theory and experiment are in excellent agreement. The magnetic field control is so accurate that it should be possible to resolve changes in the elastic scattering length on the broadest resonance at 1007 G.

The work in Garching is supported by the European-Union network ``Cold Quantum Gases", the DFG-For\-scher\-gruppe ``Quantengase", and GIF. The work at Eindhoven is part of the research program of the Stichting FOM, which is financially supported by NWO.

%\bibliographystyle{apsrev}
%\bibliography{bib2002}

\begin{thebibliography}{41}
\expandafter\ifx\csname natexlab\endcsname\relax\def\natexlab#1{#1}\fi
\expandafter\ifx\csname bibnamefont\endcsname\relax
  \def\bibnamefont#1{#1}\fi
\expandafter\ifx\csname bibfnamefont\endcsname\relax
  \def\bibfnamefont#1{#1}\fi
\expandafter\ifx\csname citenamefont\endcsname\relax
  \def\citenamefont#1{#1}\fi
\expandafter\ifx\csname url\endcsname\relax
  \def\url#1{\texttt{#1}}\fi
\expandafter\ifx\csname urlprefix\endcsname\relax\def\urlprefix{URL }\fi
\providecommand{\bibinfo}[2]{#2}
\providecommand{\eprint}[2][]{\url{#2}}

\bibitem[{\citenamefont{{Inouye \it et al.}}(1998)}]{inouye:98}
\bibinfo{author}{\bibfnamefont{S.}~\bibnamefont{{Inouye \it et al.}}},
  \bibinfo{journal}{Nature (London)} \textbf{\bibinfo{volume}{392}},
  \bibinfo{pages}{151} (\bibinfo{year}{1998}).

\bibitem[{\citenamefont{{Courteille \it et al.}}(1998)}]{courteille:98}
\bibinfo{author}{\bibfnamefont{P.}~\bibnamefont{{Courteille \it et al.}}},
  \bibinfo{journal}{Phys. Rev. Lett.} \textbf{\bibinfo{volume}{81}},
  \bibinfo{pages}{69} (\bibinfo{year}{1998}).

\bibitem[{\citenamefont{{Roberts \it et al.}}(1998)}]{roberts:98}
\bibinfo{author}{\bibfnamefont{J.}~\bibnamefont{{Roberts \it et al.}}},
  \bibinfo{journal}{Phys. Rev. Lett.} \textbf{\bibinfo{volume}{81}},
  \bibinfo{pages}{5109} (\bibinfo{year}{1998}).

\bibitem[{\citenamefont{{Vuletic \it et al.}}(1999)}]{vuletic:99}
\bibinfo{author}{\bibfnamefont{V.}~\bibnamefont{{Vuletic \it et al.}}},
  \bibinfo{journal}{Phys. Rev. Lett.} \textbf{\bibinfo{volume}{82}},
  \bibinfo{pages}{1406} (\bibinfo{year}{1999}).

\bibitem[{\citenamefont{{Loftus \it et al.}}(2002)}]{loftus:02}
\bibinfo{author}{\bibfnamefont{T.}~\bibnamefont{{Loftus \it et al.}}},
  \bibinfo{journal}{Phys. Rev. Lett.} \textbf{\bibinfo{volume}{88}},
  \bibinfo{pages}{173201} (\bibinfo{year}{2002}).

\bibitem[{\citenamefont{{Khaykovich \it et al.}}(2002)}]{khayokovich:02}
\bibinfo{author}{\bibfnamefont{L.}~\bibnamefont{{Khaykovich \it et al.}}},
  \bibinfo{journal}{Science} \textbf{\bibinfo{volume}{296}},
  \bibinfo{pages}{1290} (\bibinfo{year}{2002}).

\bibitem[{\citenamefont{{Strecker \it et al.}}(2002)}]{strecker:02}
\bibinfo{author}{\bibfnamefont{K.}~\bibnamefont{{Strecker \it et al.}}},
  \bibinfo{journal}{Nature (London)} \textbf{\bibinfo{volume}{417}},
  \bibinfo{pages}{150} (\bibinfo{year}{2002}).

\bibitem[{\citenamefont{{Donley \it et al.}}(2001)}]{donley:01}
\bibinfo{author}{\bibfnamefont{E.}~\bibnamefont{{Donley \it et al.}}},
  \bibinfo{journal}{Nature (London)} \textbf{\bibinfo{volume}{412}},
  \bibinfo{pages}{295} (\bibinfo{year}{2001}).

\bibitem[{\citenamefont{{Donley \it et al.}}(2002)}]{donley:02}
\bibinfo{author}{\bibfnamefont{E.}~\bibnamefont{{Donley \it et al.}}},
  \bibinfo{journal}{Nature (London)} \textbf{\bibinfo{volume}{417}},
  \bibinfo{pages}{529} (\bibinfo{year}{2002}).

\bibitem[{\citenamefont{{Greiner \it et al.}}(2002)}]{greiner:02}
\bibinfo{author}{\bibfnamefont{M.}~\bibnamefont{{Greiner \it et al.}}},
  \bibinfo{journal}{Nature (London)} \textbf{\bibinfo{volume}{415}},
  \bibinfo{pages}{39} (\bibinfo{year}{2002}).

\bibitem[{\citenamefont{Girardeau and Wright}(2002)}]{girardeau:01}
\bibinfo{author}{\bibfnamefont{M.}~\bibnamefont{Girardeau}} \bibnamefont{and}
  \bibinfo{author}{\bibfnamefont{E.}~\bibnamefont{Wright}},
  \bibinfo{journal}{Laser Physics} \textbf{\bibinfo{volume}{12}},
  \bibinfo{pages}{8} (\bibinfo{year}{2002}).

\bibitem[{\citenamefont{{Dalfovo \it et al.}}(1999)}]{dalfovo:99}
\bibinfo{author}{\bibfnamefont{F.}~\bibnamefont{{Dalfovo \it et al.}}},
  \bibinfo{journal}{Rev. Mod. Phys.} \textbf{\bibinfo{volume}{71}},
  \bibinfo{pages}{463} (\bibinfo{year}{1999}).

\bibitem{kokkelmans:etal}
\bibinfo{author}{\bibfnamefont{S.}~\bibnamefont{Kokkelmans}},
  \bibinfo{author}{\bibfnamefont{H.}~\bibnamefont{Visser}}, \bibnamefont{and}
  \bibinfo{author}{\bibfnamefont{B.}~\bibnamefont{Verhaar}},
  \bibinfo{journal}{Phys. Rev. A} \textbf{\bibinfo{volume}{63}},
  \bibinfo{pages}{31601} (\bibinfo{year}{2001});
\bibinfo{author}{\bibfnamefont{M.}~\bibnamefont{Mackie}}
  (\bibinfo{year}{physics/0202041}).

\bibitem{chin:etal}
\bibinfo{author}{\bibfnamefont{C.}~\bibnamefont{{Chin \it et al.}}},
  \bibinfo{journal}{Phys. Rev. Lett.} \textbf{\bibinfo{volume}{85}},
  \bibinfo{pages}{2717} (\bibinfo{year}{2000});
\bibinfo{author}{\bibfnamefont{P.}~\bibnamefont{Leo}},
  \bibinfo{author}{\bibfnamefont{C.}~\bibnamefont{Williams}}, \bibnamefont{and}
  \bibinfo{author}{\bibfnamefont{P.}~\bibnamefont{Julienne}},
  \bibinfo{journal}{{\it ibid.}} \textbf{\bibinfo{volume}{85}},
  \bibinfo{pages}{2721} (\bibinfo{year}{2000}).

\bibitem[{\citenamefont{{Stenger \it et al.}}(1999)}]{stenger:99}
\bibinfo{author}{\bibfnamefont{J.}~\bibnamefont{{Stenger \it et al.}}},
  \bibinfo{journal}{Phys. Rev. Lett.} \textbf{\bibinfo{volume}{82}},
  \bibinfo{pages}{2422} (\bibinfo{year}{1999}).

\bibitem[{\citenamefont{{Roberts \it et al.}}(2000)}]{roberts:00}
\bibinfo{author}{\bibfnamefont{J.}~\bibnamefont{{Roberts \it et al.}}},
  \bibinfo{journal}{Phys. Rev. Lett.} \textbf{\bibinfo{volume}{85}},
  \bibinfo{pages}{728} (\bibinfo{year}{2000}).

\bibitem[{\citenamefont{Newbury et~al.}(1995)\citenamefont{Newbury, Myatt, and
  Wieman}}]{newbury:95}
\bibinfo{author}{\bibfnamefont{N.}~\bibnamefont{Newbury}},
  \bibinfo{author}{\bibfnamefont{C.}~\bibnamefont{Myatt}}, \bibnamefont{and}
  \bibinfo{author}{\bibfnamefont{C.}~\bibnamefont{Wieman}},
  \bibinfo{journal}{Phys. Rev. A} \textbf{\bibinfo{volume}{51}},
  \bibinfo{pages}{2680} (\bibinfo{year}{1995}).

\bibitem[{\citenamefont{{Myatt \it et al.}}(1997)}]{myatt:97}
\bibinfo{author}{\bibfnamefont{C.}~\bibnamefont{{Myatt \it et al.}}},
  \bibinfo{journal}{Phys. Rev. Lett.} \textbf{\bibinfo{volume}{78}},
  \bibinfo{pages}{586} (\bibinfo{year}{1997}).

\bibitem[{\citenamefont{{Wynar \it et al.}}(2000)}]{wynar:00}
\bibinfo{author}{\bibfnamefont{R.}~\bibnamefont{{Wynar \it et al.}}},
  \bibinfo{journal}{Science} \textbf{\bibinfo{volume}{287}},
  \bibinfo{pages}{1016} (\bibinfo{year}{2000}).

\bibitem[{\citenamefont{{Seto \it et al.}}(2000)}]{seto:00}
\bibinfo{author}{\bibfnamefont{J.}~\bibnamefont{{Seto \it et al.}}},
  \bibinfo{journal}{J. Chem. Phys.} \textbf{\bibinfo{volume}{113}},
  \bibinfo{pages}{3067} (\bibinfo{year}{2000}).

\bibitem[{\citenamefont{{Roberts \it et al.}}(2001)}]{roberts:01a}
\bibinfo{author}{\bibfnamefont{J.}~\bibnamefont{{Roberts \it et al.}}},
  \bibinfo{journal}{Phys. Rev. A} \textbf{\bibinfo{volume}{64}},
  \bibinfo{pages}{24702} (\bibinfo{year}{2001}).

\bibitem[{\citenamefont{{Vogels \it et al.}}(1997)}]{vogels:97}
\bibinfo{author}{\bibfnamefont{J.}~\bibnamefont{{Vogels \it et al.}}},
  \bibinfo{journal}{Phys. Rev. A} \textbf{\bibinfo{volume}{56}},
  \bibinfo{pages}{1067} (\bibinfo{year}{1997}).

\bibitem[{\citenamefont{{van Kempen \it et al.}}(2002)}]{kempen:02}
\bibinfo{author}{\bibfnamefont{E.}~\bibnamefont{{van Kempen \it et al.}}},
  \bibinfo{journal}{Phys. Rev. Lett.} \textbf{\bibinfo{volume}{88}},
  \bibinfo{pages}{93201} (\bibinfo{year}{2002}).

\bibitem[{\citenamefont{{Ernst \it et al.}}(1998)}]{ernst:98}
\bibinfo{author}{\bibfnamefont{U.}~\bibnamefont{{Ernst \it et al.}}},
  \bibinfo{journal}{Europhys. Lett.} \textbf{\bibinfo{volume}{41}},
  \bibinfo{pages}{1} (\bibinfo{year}{1998}).

\bibitem[{\citenamefont{Heinzen}(1999)}]{heinzen:99}
\bibinfo{author}{\bibfnamefont{D.}~\bibnamefont{Heinzen}},   
  \emph{\bibinfo{title}{Ultracold Atomic Interactions}}, in
  \emph{\bibinfo{booktitle}{Proceedings of the International School of Physics {``Enrico
  Fermi"}, Course CXL}}, edited by
  \bibinfo{editor}{\bibfnamefont{M.}~\bibnamefont{{Inguscio \it et al.}}}
  (\bibinfo{publisher}{IOS Press}, \bibinfo{address}{Amsterdam},
  \bibinfo{year}{1999}), p. \bibinfo{pages}{351}.

\bibitem[{\citenamefont{Burke}(1999)}]{burke:99}
\bibinfo{author}{\bibfnamefont{J.}~\bibnamefont{Burke}}, Ph.D. thesis,
  \bibinfo{school}{University of Colorado, 1999}, available at \\
  http://amo.phy.gasou.edu/bec.html/bibliography.html.

\bibitem[{\citenamefont{Mizushima}(1975)}]{mizushima:etal}
\bibinfo{author}{\bibfnamefont{M.}~\bibnamefont{Mizushima}},
  \emph{\bibinfo{title}{The theory of rotating diatomic molecules}}
  (\bibinfo{publisher}{Wiley}, \bibinfo{address}{New York},
  \bibinfo{year}{1975}), \bibinfo{note}{p.\ 223};
\bibinfo{author}{\bibfnamefont{F.}~\bibnamefont{{Mies \it et al.}}},
  \bibinfo{journal}{J.\ Res.\ Natl.\ Inst.\ Stand.\ Technol.}
  \textbf{\bibinfo{volume}{101}}, \bibinfo{pages}{521} (\bibinfo{year}{1996}).

\bibitem[{\citenamefont{Marinescu et~al.}(1994)\citenamefont{Marinescu,
  Sadeghpour, and Dalgarno}}]{marinescu:94}
\bibinfo{author}{\bibfnamefont{M.}~\bibnamefont{Marinescu}},
  \bibinfo{author}{\bibfnamefont{H.}~\bibnamefont{Sadeghpour}},
  \bibnamefont{and} \bibinfo{author}{\bibfnamefont{A.}~\bibnamefont{Dalgarno}},
  \bibinfo{journal}{Phys. Rev. A} \textbf{\bibinfo{volume}{49}},
  \bibinfo{pages}{982} (\bibinfo{year}{1994}).

\bibitem[{\citenamefont{Fedichev et~al.}(1996)\citenamefont{Fedichev, Reynolds,
  and Shlyapnikov}}]{fedichev:etal}
\bibinfo{author}{\bibfnamefont{P.}~\bibnamefont{Fedichev}},
  \bibinfo{author}{\bibfnamefont{M.}~\bibnamefont{Reynolds}}, \bibnamefont{and}
  \bibinfo{author}{\bibfnamefont{G.}~\bibnamefont{Shlyapnikov}},
  \bibinfo{journal}{Phys. Rev. Lett.} \textbf{\bibinfo{volume}{77}},
  \bibinfo{pages}{2921} (\bibinfo{year}{1996});
\bibinfo{author}{\bibfnamefont{F.}~\bibnamefont{van Abeelen}} \bibnamefont{and}
  \bibinfo{author}{\bibfnamefont{B.}~\bibnamefont{Verhaar}},
  \bibinfo{journal}{{\it ibid.}} \textbf{\bibinfo{volume}{83}},
  \bibinfo{pages}{1550} (\bibinfo{year}{1999});
\bibinfo{author}{\bibfnamefont{E.}~\bibnamefont{Nielsen}} \bibnamefont{and}
  \bibinfo{author}{\bibfnamefont{J.}~\bibnamefont{Macek}},
  \bibinfo{journal}{{\it ibid.}} \textbf{\bibinfo{volume}{83}},
  \bibinfo{pages}{1566} (\bibinfo{year}{1999});
\bibinfo{author}{\bibfnamefont{B.}~\bibnamefont{Esry}},
  \bibinfo{author}{\bibfnamefont{C.}~\bibnamefont{Greene}}, \bibnamefont{and}
  \bibinfo{author}{\bibfnamefont{J.}~\bibnamefont{Burke}},
  \bibinfo{journal}{{\it ibid.}} \textbf{\bibinfo{volume}{83}},
  \bibinfo{pages}{1751} (\bibinfo{year}{1999});
\bibinfo{author}{\bibfnamefont{E.}~\bibnamefont{Braaten}} \bibnamefont{and}
  \bibinfo{author}{\bibfnamefont{H.}~\bibnamefont{Hammer}},
  \bibinfo{journal}{{\it ibid.}} \textbf{\bibinfo{volume}{87}},
  \bibinfo{pages}{160407} (\bibinfo{year}{2001}).

\bibitem[{\citenamefont{{Burt \it et al.}}(1997)}]{burt:97}
\bibinfo{author}{\bibfnamefont{E.}~\bibnamefont{{Burt \it et al.}}},
  \bibinfo{journal}{Phys. Rev. Lett.} \textbf{\bibinfo{volume}{79}},
  \bibinfo{pages}{337} (\bibinfo{year}{1997}).

\bibitem[{\citenamefont{{Claussen \it et al.}}(2002)}]{claussen:etal}
\bibinfo{author}{\bibfnamefont{N.}~\bibnamefont{{Claussen \it et al.}}},
  \bibinfo{journal}{Phys. Rev. Lett.} \textbf{\bibinfo{volume}{89}},
  \bibinfo{pages}{10401} (\bibinfo{year}{2002});
\bibinfo{author}{\bibfnamefont{S.}~\bibnamefont{Kokkelmans}} \bibnamefont{and}
  \bibinfo{author}{\bibfnamefont{M.}~\bibnamefont{Holland}}
  (\bibinfo{year}{cond-mat/0204504});
\bibinfo{author}{\bibfnamefont{R.}~\bibnamefont{Duine}} \bibnamefont{and}
  \bibinfo{author}{\bibfnamefont{H.}~\bibnamefont{Stoof}}
  (\bibinfo{year}{cond-mat/0204529});
\bibinfo{author}{\bibfnamefont{M.}~\bibnamefont{Mackie}},
  \bibinfo{author}{\bibfnamefont{K.}~\bibnamefont{Suominen}}, \bibnamefont{and}
  \bibinfo{author}{\bibfnamefont{J.}~\bibnamefont{Javanainen}}
  (\bibinfo{year}{cond-mat/0205535}).

\end{thebibliography}

\end{document}